 \documentclass[preprint2]{aastex}

\usepackage{natbib}
\newcommand{\nht}{$N_{\rm H_2}$}
\def\degr{\hbox{$^\circ$}}

\def\arcsec{\hbox{$^{\prime\prime}$}}
\newcommand{\microns}{$\mu$m }

\shorttitle{The structure of infrared dark clouds}
\shortauthors{Peretto \& Fuller}

\begin{document}


\title{A statistical study of the mass and density structure of Infrared Dark Clouds}


\author{N. Peretto$^{1,2}$ and G. A. Fuller$^{1}$}
\affil{1. Jodrell Bank Centre for Astrophysics, School of Physics \& Astronomy, 
University of Manchester, Oxford Road, Manchester M13 9PL, UK\\
2. Laboratoire AIM, CEA/DSM-CNRS-Universit\'e Paris Diderot, IRFU/Service d'Astrophysique, C.E. Saclay, Orme des merisiers, 91191 Gif-sur-Yvette, France}

\email{nicolas.peretto@manchester.ac.uk}

\begin{abstract}
  How and when the mass distribution of stars in the Galaxy is set is
  one of the main issues of modern astronomy. Here we present a
  statistical study of mass and density distributions of infrared dark
  clouds (IRDCs) and fragments within them.  These regions are
  pristine molecular gas structures and progenitors of stars and so
  provide insights into the initial conditions of star formation.
  This study makes use of a IRDC catalogue (Peretto \& Fuller 2009),
  the largest sample of IRDC column density maps to date, containing a
  total of $\sim$11,000 IRDCs with column densities exceeding $N_{H_2}
  = 1\times10^{22}$~cm$^{-2}$ and over 50,000 single peaked IRDC
  fragments.  The large number of objects constitutes an important
  strength of this study, allowing detailed analysis of the
  completeness of the sample and so statistically robust conclusions.
  Using a statistical approach to assigning distances to clouds, the
  mass and density distributions of the clouds and the fragments
  within them are constructed.  The mass distributions show a
  steepening of the slope when switching from IRDCs to fragments, in
  agreement with previous results of similar structures.  IRDCs and 
  fragments are divided into unbound/bound objects by assuming 
  Larson's relation and calculating their virial parameter. IRDCs are mostly gravitationally bound, while a significant fraction of the fragments are not. The density distribution of
  gravitationally unbound fragments shows a steep characteristic slope
  such as $\Delta N/ \Delta \log(n) \propto n^{-4.0 \pm 0.5}$, rather independent of
  the range of fragment mass. However, the incompleteness limit at a
  number density of $\sim10^3$~cm$^{-3}$ does not allow us to exclude
  a potential lognormal density distribution.  In contrast,
  gravitationally bound fragments show a characteristic density peak
  at $n\simeq 10^4$~cm$^{-3}$ but the shape of the density
  distributions changes with the range of fragment masses.  An
  explanation for this could be differential dynamical evolution of
  the fragment density with respect to their mass as more massive
  fragments contract more rapidly.  The IRDC properties reported here
  provide a representative view of the density and mass structure of dense molecular clouds before and during the earliest stages of star formation. These should serve as constraints on any theoretical or numerical model to identify the physical processes involved in the formation and evolution of
  structure in molecular clouds.
\end{abstract}


\section{Introduction}

While low mass stars dominate the mass of galaxies, massive stars regulate
their energy budget. Understanding how and when the mass distribution of stars
is determined is therefore essential in establishing a comprehensive picture
of galactic evolution, and star formation, throughout the Universe.

\begin{figure*}[!th!]
\hspace{0cm}
\includegraphics[width=16.5cm,angle=0]{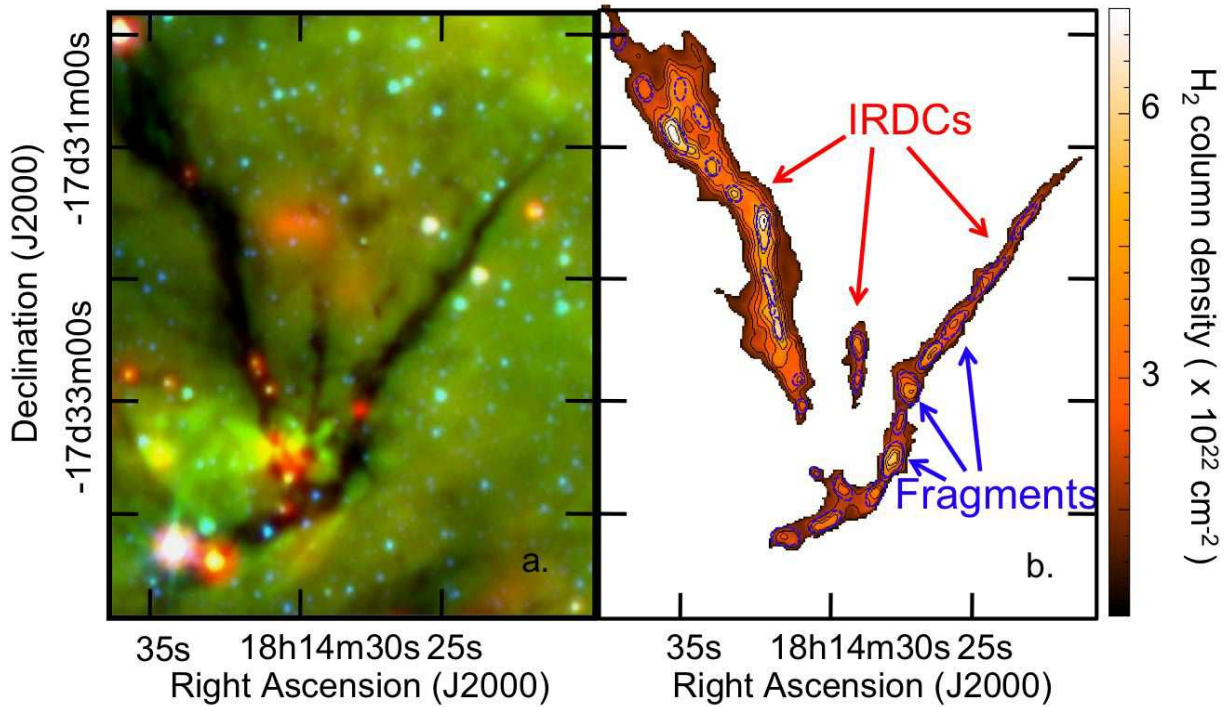}
\caption{Filamentary infrared dark clouds. ({\it a}) Spitzer 3 colour image of
  a region containing IRDCs (Blue: 3.6\microns; Green: 8\microns; Red:
  24\microns). The IRDCs are the long filamentary structures seen in
  extinction. On this figure the blue stars are foreground stars, the
  red/yellow stars are young stars currently forming in the IRDC. ({\it b})
  H$_2$ column density map constructed from the 8\microns extinction seen in
  ({\it a}). While the outer contour delimits the boundary of each of the
  three IRDCs, the fragments are the substructures seen within each IRDCs. The
  28 fragments identified in these clouds are marked by blue dashed ellipses.}
\end{figure*}

Since stars form in molecular clouds the comparison of the internal structure
of the clouds and the initial mass function (IMF) of stars can provide
insights on the processes responsible for the formation of stars.  The mass
distribution of molecular clouds, and cores within them have been extensively
studied in the past twenty years. Until recently, it was believed that the
mass distribution of CO clumps was described by $\Delta N_{\rm {CO}}/\Delta
\log M=M^{-\alpha}$ with $\alpha=0.7\pm 0.2$ for the Milky Way (Kramer et
al. 1998, Rosolowski 2004). The mass distribution of prestellar cores, the
direct progenitors of stars and stellar systems, as observed in dust continuum
is much steeper, resembling the Salpeter IMF with a power law index of
$\alpha=1.35$ (Motte et al. 1998; Enoch et al. 2008). However, recent papers
questioned the impact of the source extraction scheme used to segment the data
on the final mass distribution shape (Pineda et al. 2009). Buckle et
al. (2010) found a steeper mass distribution for small scale CO clumps. Also,
in most cases, different tracers are required to trace different structures
such as dense cores and molecular clumps, raising the question of detection
biases. Statistics is often a problem too, binning small number of objects
introduce artifacts (Reid \& Wilson 2006). Therefore some confusion exists on
what is the real mass structure of molecular clouds.

Another important physical aspect of molecular cloud structure is the
probability density function (PDF) of the gas volume density. This quantity
has received only little attention (e.g. Dring et al. 1996 for HI; Smith
\& Scalo 2009 for CO) but potentially contains crucial
information on the processes at the origin of the density fluctuations.
 For instance, turbulence-driven fragmentation models develop
initial lognormal density fluctuations (e.g. Padoan et al. 1997), which could
be the main driver of the lognormal part of the IMF (Chabrier 2003). Studying
the density distribution of fragments within molecular clouds could set
important constraints on such models.

To perform such studies, we decided to focus on a specific type of molecular clouds, i.e. 
infrared dark clouds (IRDCs). IRDCs are dense molecular clouds seen in silhouette
against the bright emission of the galactic plane (e.g. Perault et al. 1996;
Teyssier et al. 2002; Rathborne et al. 2006; Simon et al. 2006).  They are
cold and only slightly processed by star formation activity, still containing
the initial conditions of star formation.  Peretto \& Fuller (2009; hereafter
PF09) recently constructed the column density maps of more than 11,000 of such
IRDCs, the largest database of such structures to date.  This catalogue
provides the opportunity to probe molecular clouds in the Galaxy over a wide
range of size scales and column density at high angular resolution using the
8\microns dust absorption and a new source extraction scheme.


In section 2 of the present paper we discuss the dataset we used. In section 3
we describe and re-analyze previous results on the distance distribution of
IRDCs, while in Section 4 we estimate completeness limits. Section 5 displays
our main results on the size, density and mass distributions of IRDCs and
fragments. Discussion is in Section 6. And finally we summarize the
main findings of this paper in Section 7.

\section{Data set}

The analysed IRDCs come from a new catalogue of clouds identified in the
Spitzer GLIMPSE data (PF09).  IRDCs were defined as connected structures with
column density peaks above $N_{\rm H_2}=2\times10^{22}$ cm$^{-2}$ and
boundaries defined by the contour at $N_{\rm H_2}=1\times10^{22}$ cm$^{-2}$.
Single peaked structures lying within the IRDCs were identified as
fragments. The boundary of a fragment being defined by the contour of the
local minimum between a fragment and its closest neighbour, the same criterion
used to define the {\it leaves} of the dendogram analysis of Rosolowsky et
al. (2008). As column density peaks, these fragments are particularly
important in the context of star formation since they are the likely birth
place of the future generation of stars.
The catalogue includes opacity maps at 4\arcsec\ resolution and physical
properties for over 11,000 IRDCs. Extracting the densest structures, a
total of $\sim$50,000 fragments have been catalogued within the full sample of
clouds (PF09).  Figure 1a shows a Spitzer three colour image of a region
containing three filamentary IRDCs from the PF09 catalogue.  Figure 1b shows
the column density map of these IRDCs and identifies the fragments within the
clouds.

%
%

\begin{figure}[!t!]
\hspace{-0.5cm}
\includegraphics[width=6cm,angle=270]{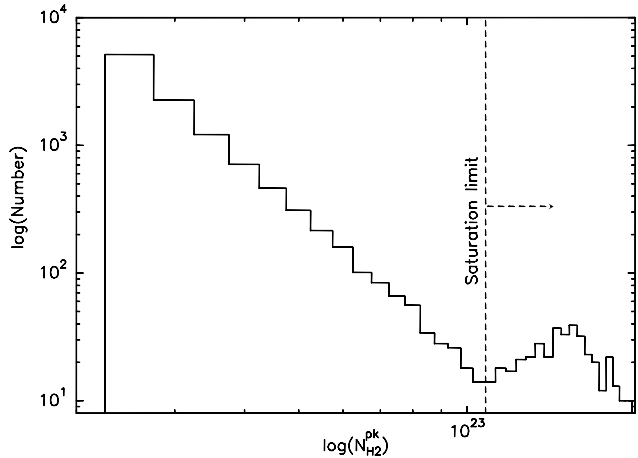}
\caption{Histogram of the IRDC peak column density. We can see a steady decrease down to 
$N_{\rm H_2} \simeq 1\times10^{23}$~cm$^{-2}$. Above this limit saturation does not allow us to probe the true peak column density. 
  \label{NH2_pk_av}}
\end{figure}

\subsection{IRDC saturation}

As discussed in PF09, some of the absorption towards IRDCs is saturated,
meaning the infrared background is not strong enough in order to fully probe
the internal structure of an IRDC. Based on photometric noise limitation and
background strength PF09 estimated the fraction of saturated IRDCs to be 3\%,
corresponding to roughly 340 IRDCs over the entire sample. 

An alternative estimate of the number of saturated clouds can be 
made from an inspection of the distribution of peak column density of IRDCs
shown in Fig.~\ref{NH2_pk_av}.
There appears to be a break in the distribution of peak \nht\ at $\sim
1\times10^{23}$~cm$^{-2}$, which likely reflects the effect of saturation.
The fraction of IRDCs lying above this limit is 4\%, very similar to the value
estimated in PF09. However, even for these saturated clouds only a small
fraction of their area is above the saturation limit, only marginally affecting
the averaged IRDC column density (and therefore any estimate of the cloud
mass). But, the saturation has a much stronger effect on some fragments.  For
this reason, in the analysis presented here IRDCs containing saturated pixels
are considered, but fragments with saturated pixels are excluded.

Our estimated saturation limit is roughly twice as large as the one found by
Vasyunina et al. (2009) from millimetre emission in their study of
particularly high column density IRDCs.  
The discrepancy between the low absorption column densities Vasyunina et
al. determined by assuming the minimum possible foreground emission, that due
to the zodical light, and the high values they determined from millimeter dust
continuum led Vasyunina et al.  to derive a relatively low saturation
limit. However, the majority of their clouds do not in fact appear saturated
as considerable substructure can be seen in the 8$\mu$m extinction maps.

\subsection{Column density and angular size distributions}

This study aims to statistically analyze the density and mass distributions of
IRDCs and their fragments. To derive such quantities we first need to know the
angular size and column density distributions as measured on the column
density maps constructed by PF09.  

\begin{figure}[!t!]
\hspace{-0.5cm}
\includegraphics[width=4cm,angle=270]{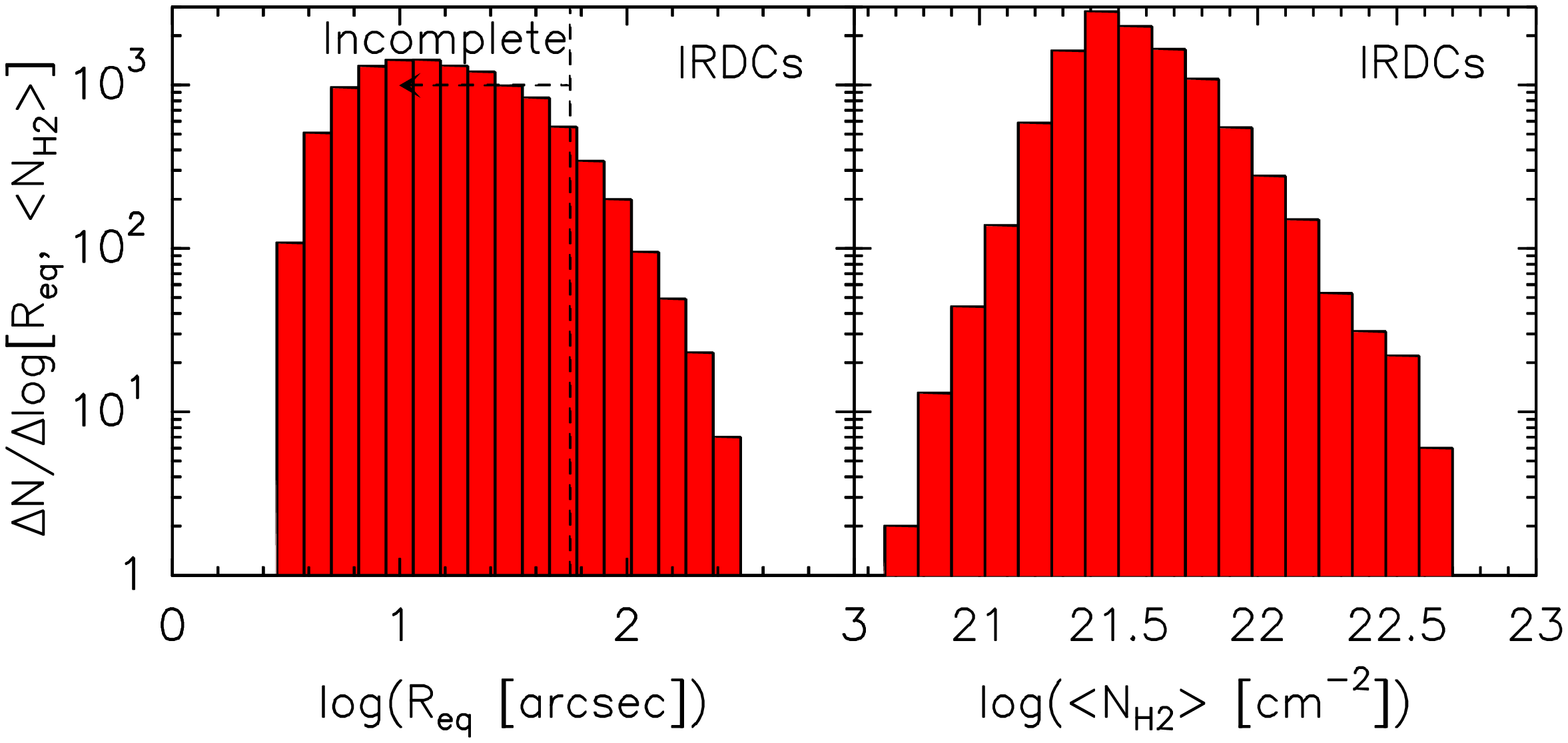}

\includegraphics[width=4cm,angle=270]{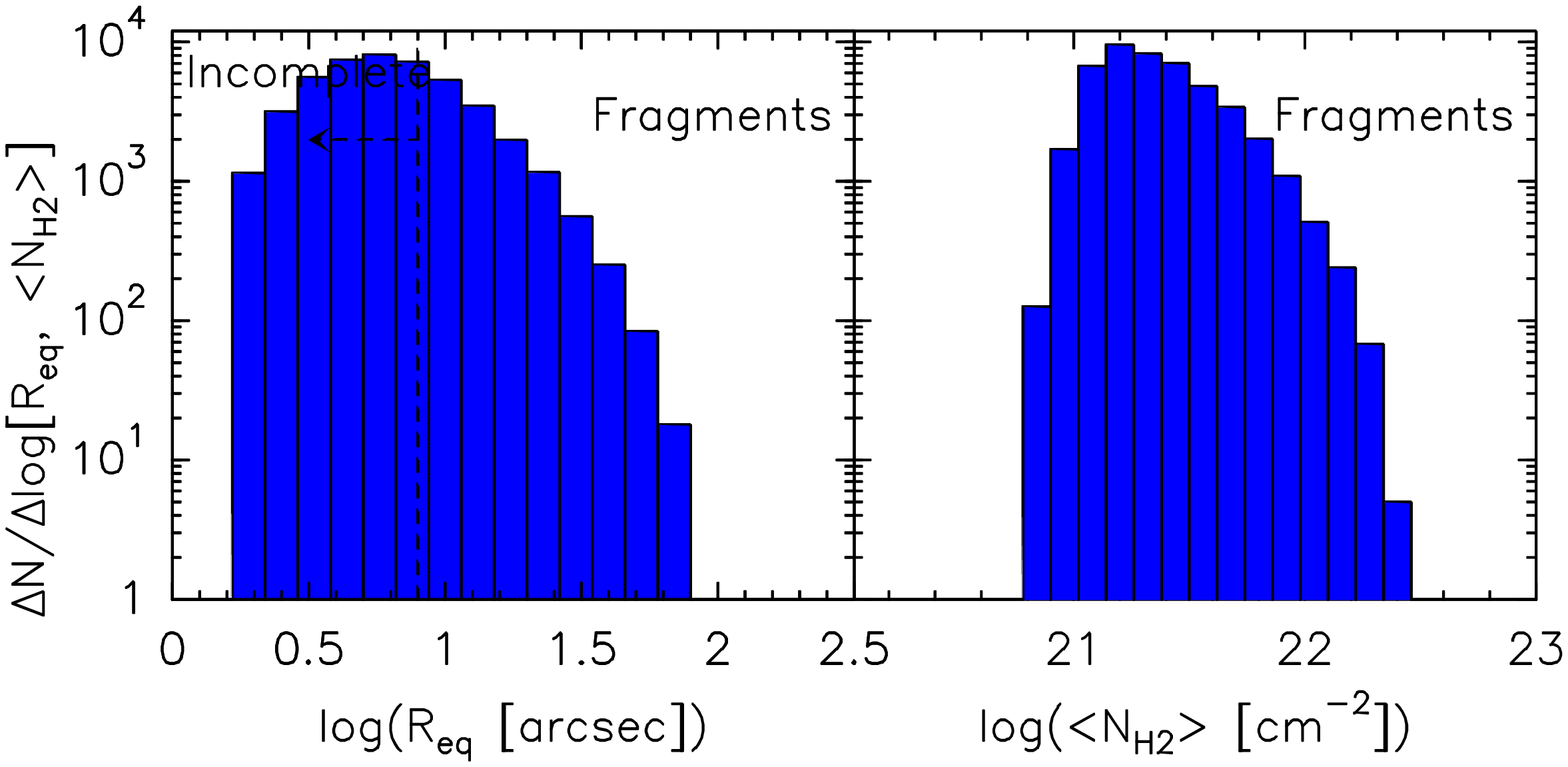}
\caption{Distributions of angular radius R$_{eq}$ and average column densities
  over each structure for IRDC (top) and fragments (bottom). The estimates for
  the radius completeness limit are discussed in Section~\ref{sec:complete}.
  \label{NH2_distrib}}
\end{figure}

Figure \ref{NH2_distrib} shows the distribution of the angular size
and column density for the $\sim11,000$ IRDCs and the $\sim 50,000$
fragments identified within them. Fragments with saturated pixels have
been excluded (see Sec.~2.1). In addition IRDCs which are not
fragmented
($\sim40\%$ of the IRDC sample) 
have also been removed to maintain a clear definition of
a fragment as a substructure within a cloud.  However, in practice
keeping these single peak clouds has little effect on the results.


It is important to note that the column densities we plot here are the {\it
  background substracted} column densities, equivalent to the one obtained in
the {\it clipping} option of the dendogram analysis of Rosolowsky et
al. (2008). In the context of centrally concentrated structures, these column
densities are the relevant  ones when interested in the physical properties of
the gas enclosed in a given radius.  Figure~\ref{NH2_distrib} clearly shows that
the distributions are dominated by small structures of low column density. We
can also clearly see the effect of incompleteness on the distributions with
the decrease in the number of sources at low radius/column density,
responsible for the formation of artificial peaks. The incompleteness in the
sample and these distributions are discussed in Section~\ref{sec:complete}.

\begin{figure*}[!t!]
\hspace{-0.0cm}
\includegraphics[width=7cm,angle=270]{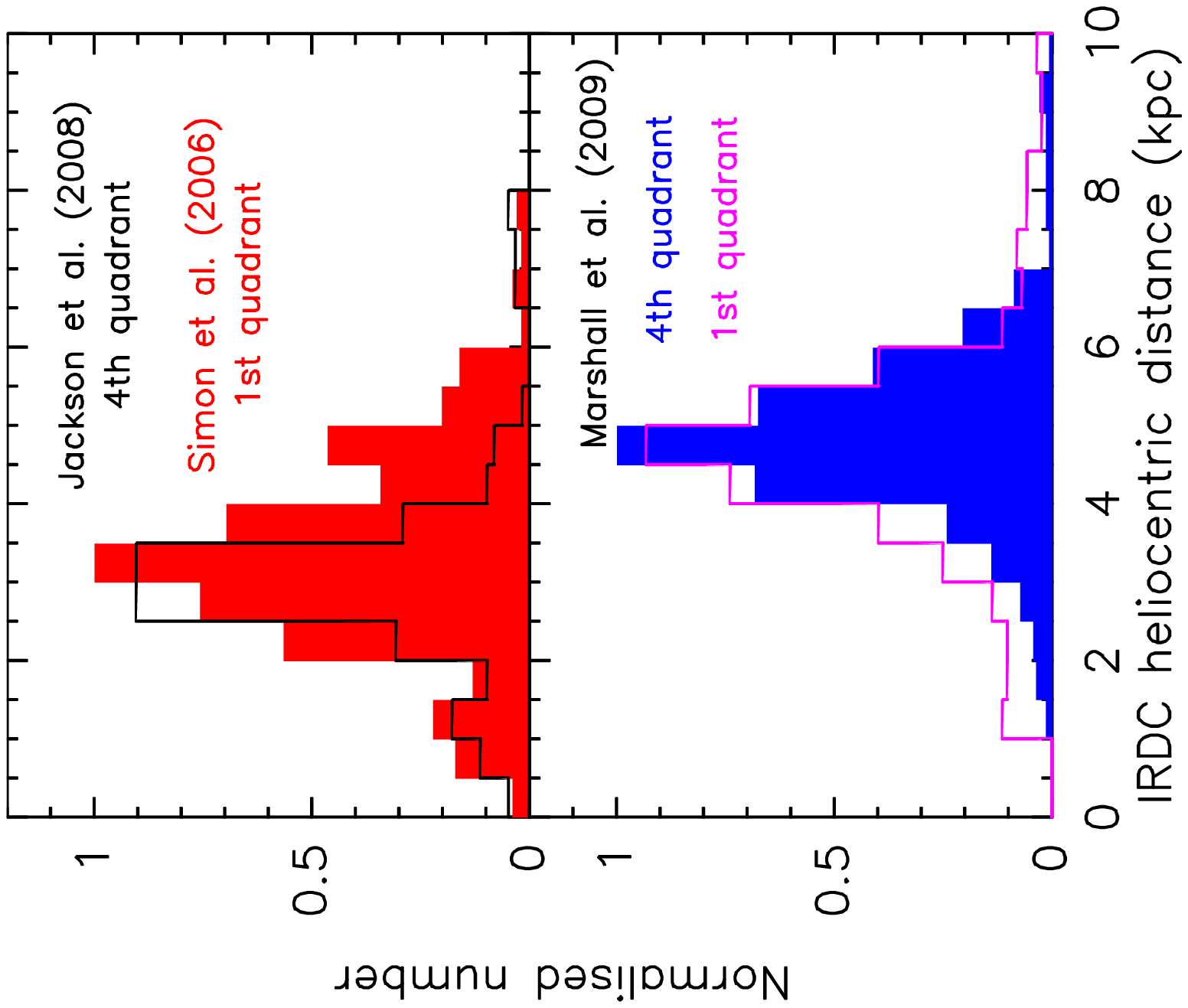}
\includegraphics[width=7cm,angle=270]{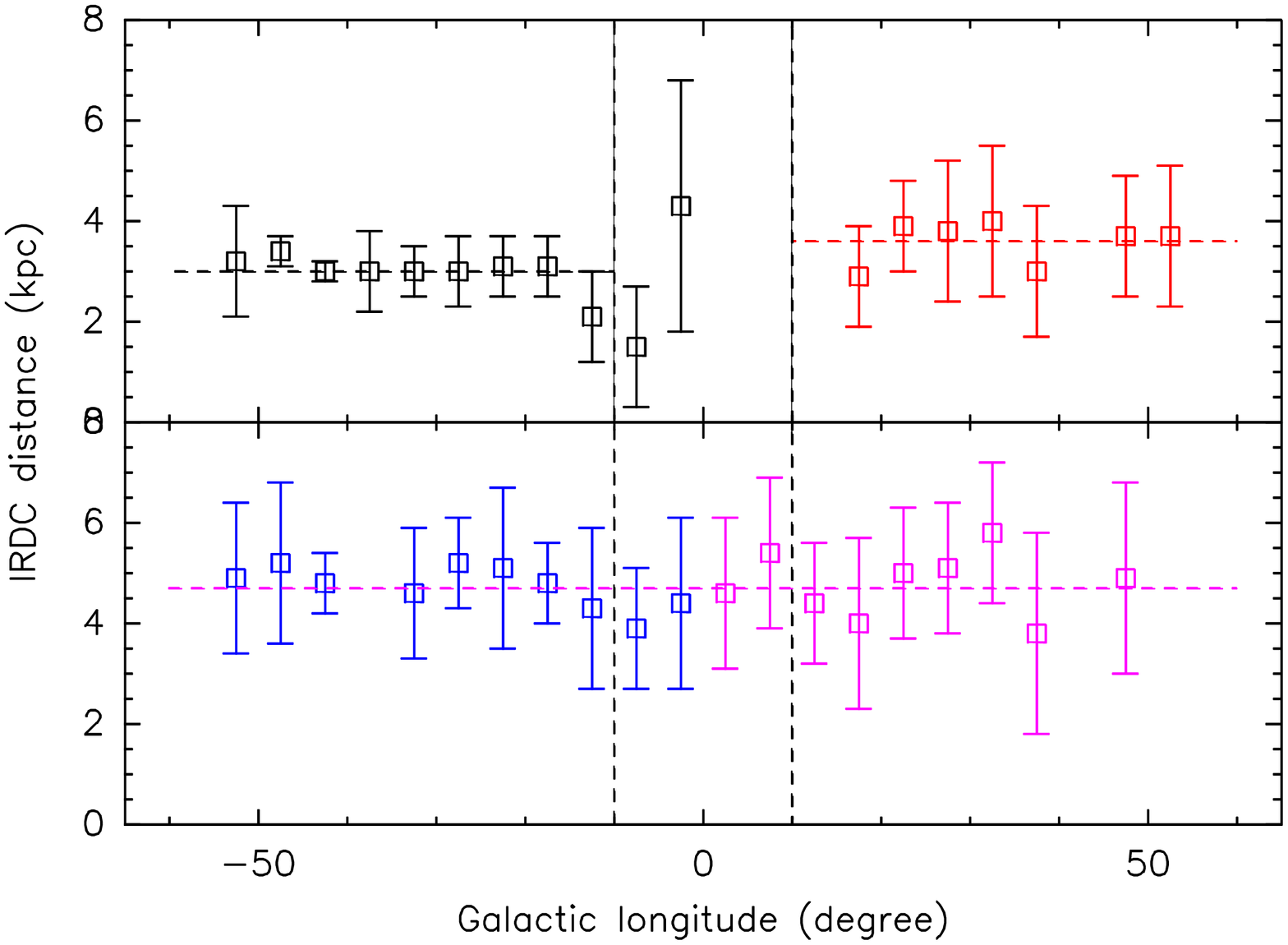}
\caption{{\it (left):} Distributions of distances of IRDCs (Simon et al. 2006, 
	Jackson et al. 2008, Marshall et al. 2009).  {\it (right):} Plot of the galactic longitude
  dependance of average IRDC distance determined from kinematics (top panel)
  and extinction (bottom panel). The horizontal dashed lines show the average
  distances over the indicated longitude range (Table~\ref{tab:dist}).
 \label{dist}}
\end{figure*}

\section{IRDCs distance distribution}

To calculate the density and mass of the clouds the distance of each IRDC is
required, however this is not yet known for most of the 11,000 IRDCs. For this
analysis we have therefore adopted a statistical approach based on previous
measurements of the distances to samples of IRDCs.

Several studies have measured the distance distribution of subsamples of IRDCs
in both the 1st and 4th quadrant of the Galactic plane
(Simon et al. 2006; Jackson et al. 2008; Marshall et al.  2009).  Both kinematic and dust
extinction techniques have been used to infer these distances, and although
they lead to similar results, there are some differences
(Fig.~\ref{dist}). The properties of these distance distributions are
summarised in Table~\ref{tab:dist}. Using dust extinction, Marshall et
al. (2009) found a centrally peaked, Gaussian-like distance distribution very
similar for both the 1st and 4th quadrant of the Galaxy (Fig.~\ref{dist} -
bottom panel).  In a similar way, kinematic distances\footnote{These kinematic
  distances have been recalculated by taking the CS(2-1) and $^{13}$CO(1-0)
  velocities published in Jackson et al. (2008) and Simon et al. (2006),
  respectively, and using the Reid et al. (2009) revised galactic rotation
  model.} shown in Fig.~\ref{dist} (top panel) show a good agreement between
the 1st and 4th quadrant . Although in the 1st quadrant a tail at 5~kpc
clearly emerges.  However most significant difference between the extinction
and kinematic distances is the position of the peak, being located at
$\sim3$~kpc in one case and at $\sim5$~kpc in the other.  Both techniques have
their own biases and advantages, it is therefore difficult to favor one
distance distribution over another. However a Gaussian distribution is a
rather good approximation to the distance distribution in both quadrants.

\begin{table*}
  \begin{tabular}{lccc}
Sample & Mean distance & FWHM & Dispersion, $\sigma$\\
       & (kpc) & (kpc) & (kpc) \\    \hline
Marshall, 1st \& 4th quad. & 4.7 & 1.5 & 1.2 \\
Jackson, 4th quad. & 3.0 & 0.8 & 1.2\\
Simon,1st quad. & 3.6 & 2.5 & 1.3\\
\hline\hline
  \end{tabular}
  \caption{Properties of the distance distributions of IRDCs shown in
    Fig.~\ref{dist}. If the distributions were truly Gaussian the FWHM would
    equal 2.35$\sigma$.}\label{tab:dist}
\end{table*}

Fig.~\ref{dist} (right) shows the average distance to the IRDCs as a function
of galactic longitude for the sources with distances measured by these two
techniques. Any systematic trend with longitude could introduce a bias in
analysis adopting a statistical distribution for the distance of IRDCs.  It is
clear that there is very little variation of the IRDC distances with respect
to the galactic longitude. The only region where there may be such an effect
is towards galactic centre seems, an area which is not covered in our IRDC
Spitzer catalogue which only extends into $l=\pm10\degr$.  It is worth noting
that distance variations around the mean as a function of longitude are very
similar for both methods, emphasising that is it predominantly only the
average distance which differs between the two methods.

\begin{figure}[!th!]
\hspace{-0.5cm}
\includegraphics[width=6cm,angle=270]{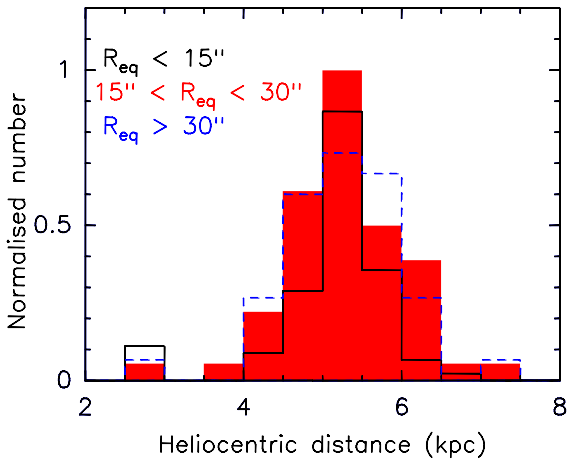}
\caption{Heliocentric distance distribution of IRDCs observed in CS J=1-0
  (Peretto et al. in prep). The 3 histograms correspond to 3 different range
  of IRDC size. There is no evidence of any statistical difference in IRDC distance as a function of  their size. 
  \label{cs_dist}}
\end{figure}

Another possible bias is with respect to the size of the IRDCs.  The studies
shown in Fig.~\ref{dist} do not contain IRDCs as small as those in the Spitzer
based sample and so it is possible that the small and large clouds have
different distance distributions.  
However, recent observations with ATNF Mopra in CS J=1-0 (Peretto et al. in
prep) of a square degree of the galactic plane ($29.8\degr < l < 31.8\degr$,
$-0.27\degr< b < 0.27\degr$) can be used to investigate this possibility.
This area covers 196 Spitzer IRDCs in total
and more than 80$\%$ of the clouds can be associated
with CS emission.

The distances of these clouds can be calculated using the Reid et al. (2009)
galactic rotation model.  Figure \ref{cs_dist} shows the distance distribution
of the Spitzer IRDCs detected in CS.  In this figure, the IRDCs have been
divided int to three size ranges. The 80 smallest IRDCs, those with
$R_{eq}$ (PF09) less than 15\arcsec~have a mean distance and standard deviation of
$5.2$~kpc and $0.8$~kpc. This is indistinguishable from the values of $5.3$~kpc
and $0.7$~kpc and $5.3$~kpc and $0.8$~kpc for the IRDCs in the next two size
ranges, $15''<R_{eq}<30''$, $R_{eq}>30''$, which contain $52$ and $39$ objects
respectively. These distributions therefore show no
indication that large and small IRDCs have different distributions of
distance.

\begin{figure*}[!t!]
\hspace{0.5cm}
\includegraphics[width=5cm,angle=270]{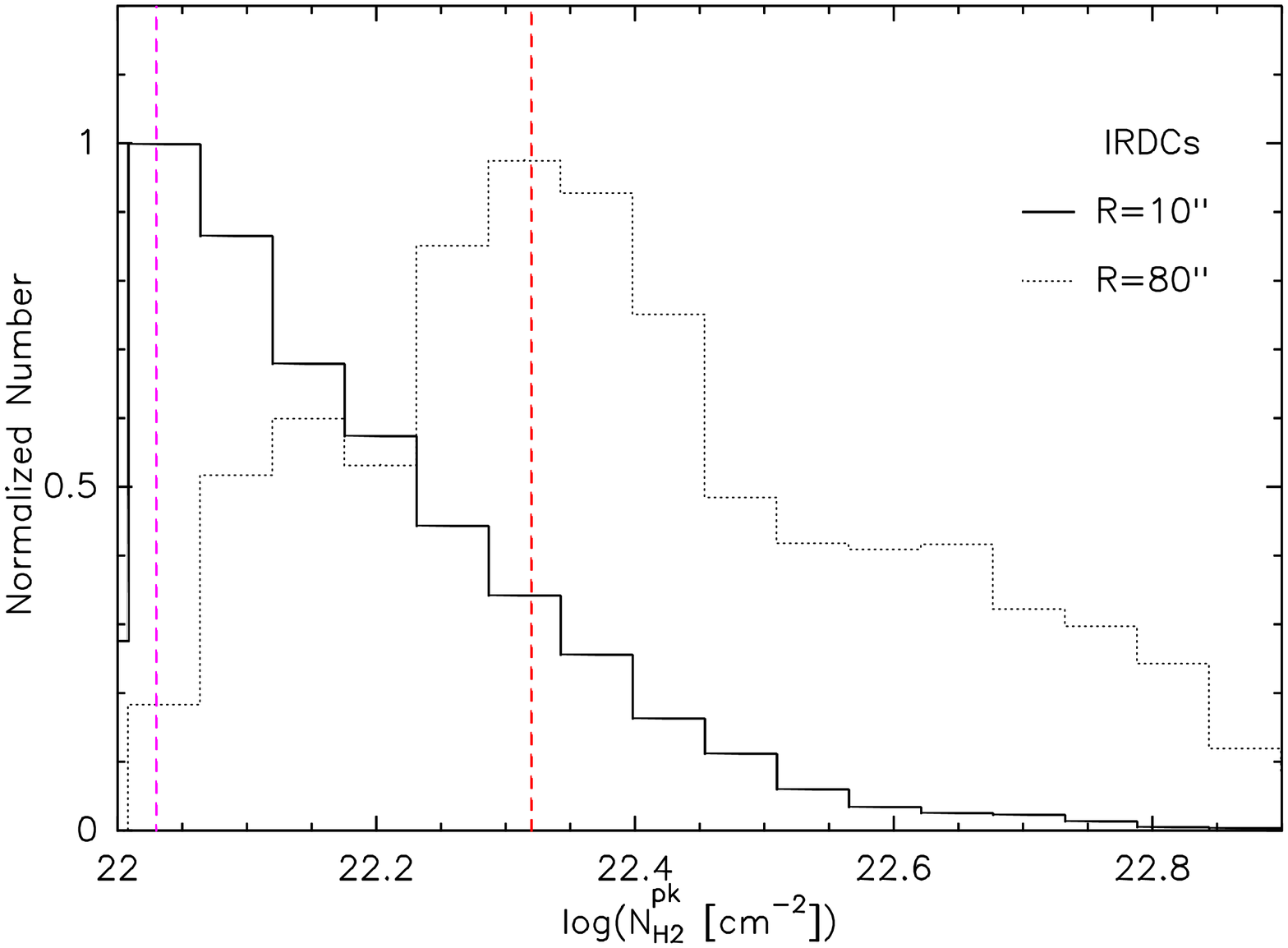}
\includegraphics[width=5cm,angle=270]{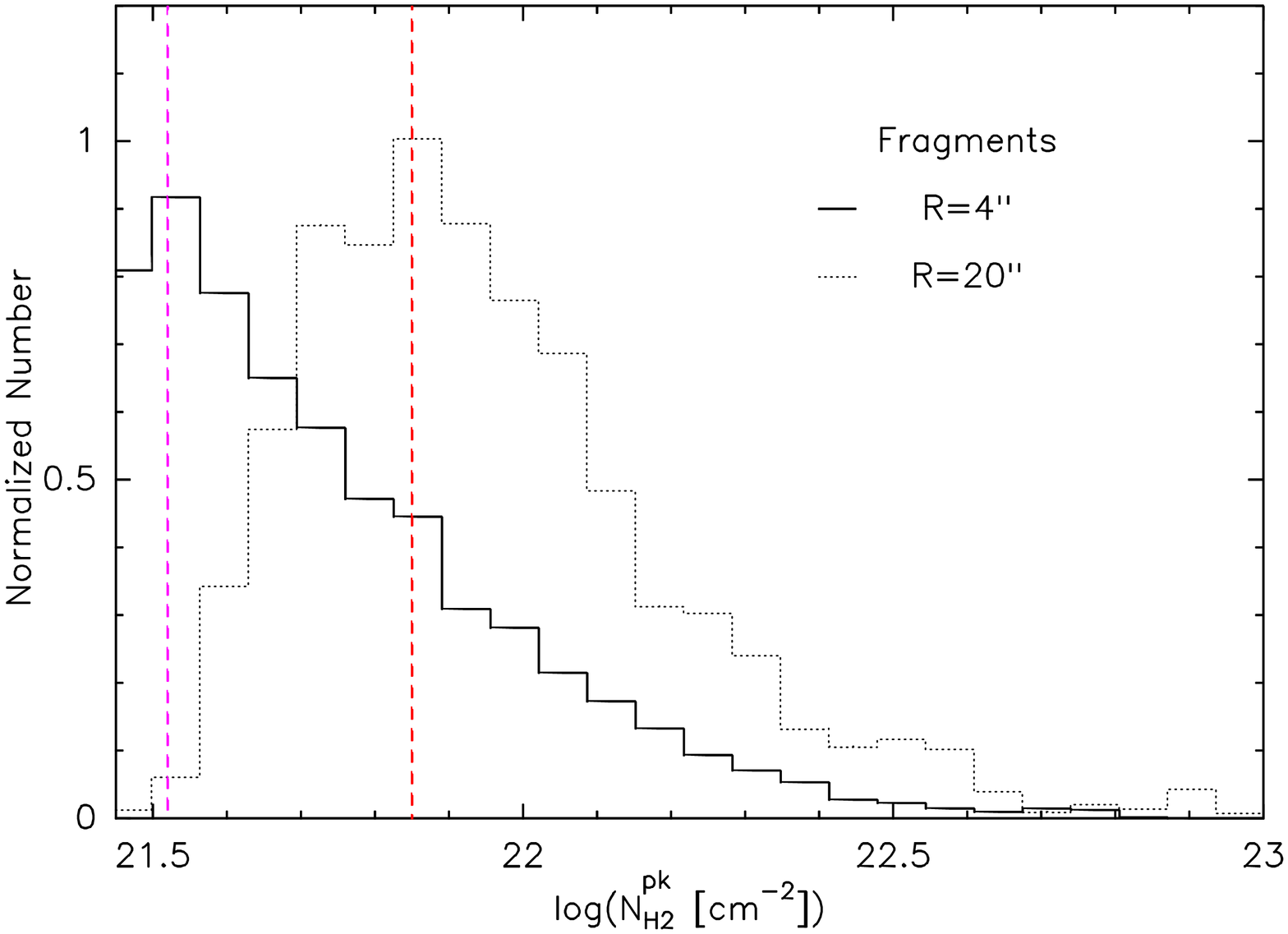}
\caption{Peak column density distributions for different IRDC (left) and
  fagment(right) sizes. For both structures we show examples for which we are
  sensitivity limited and for which we are not. The location of the peaks of
  these distributions as a function of size of the clouds and fragments are
  used to construct Fig.~\ref{rad_comp}.  }  \label{pk_r}
\end{figure*}

\begin{figure*}[!th!]
\hspace{0.0cm}
\includegraphics[width=5cm,angle=270]{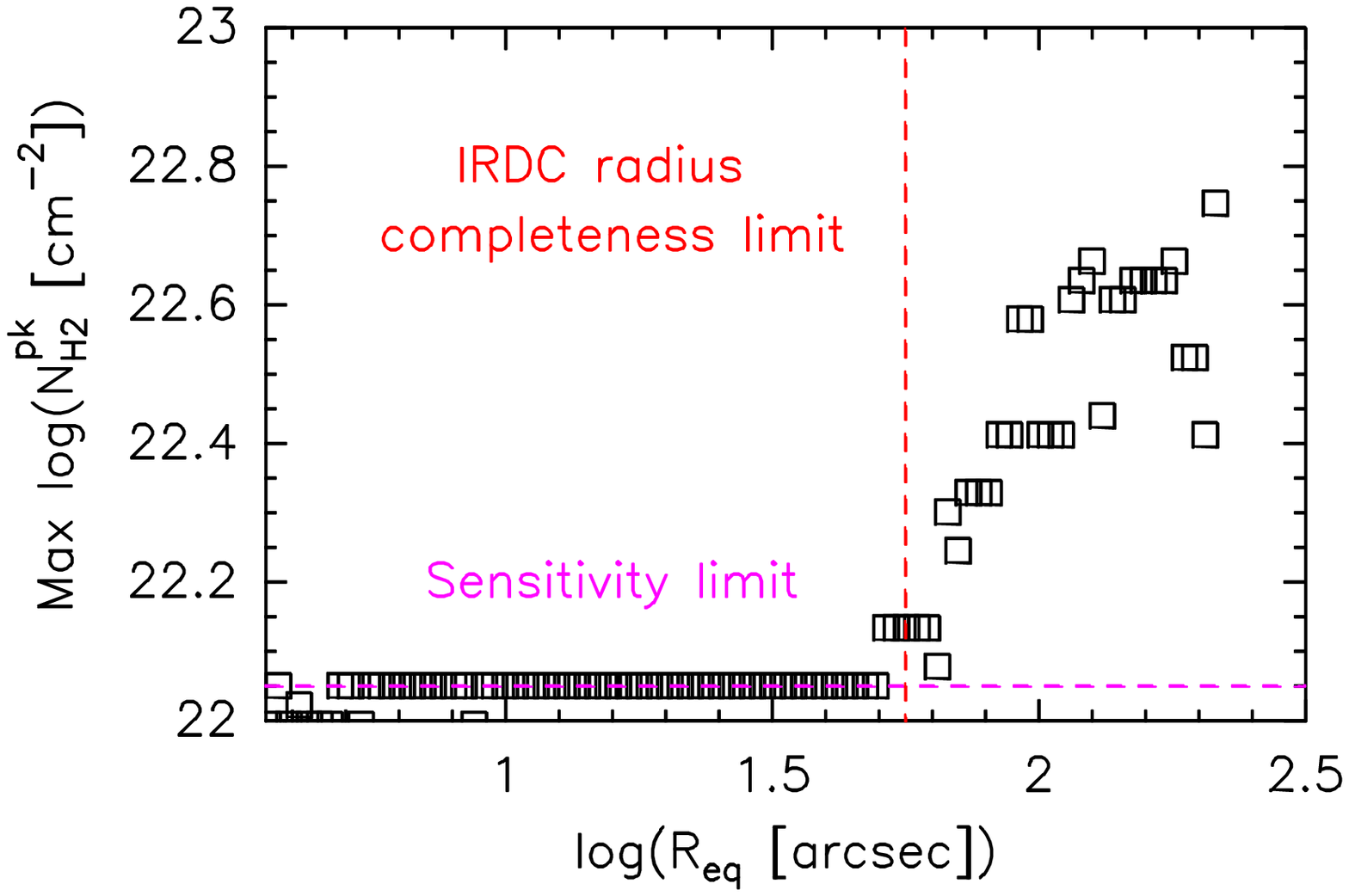}
\includegraphics[width=5cm,angle=270]{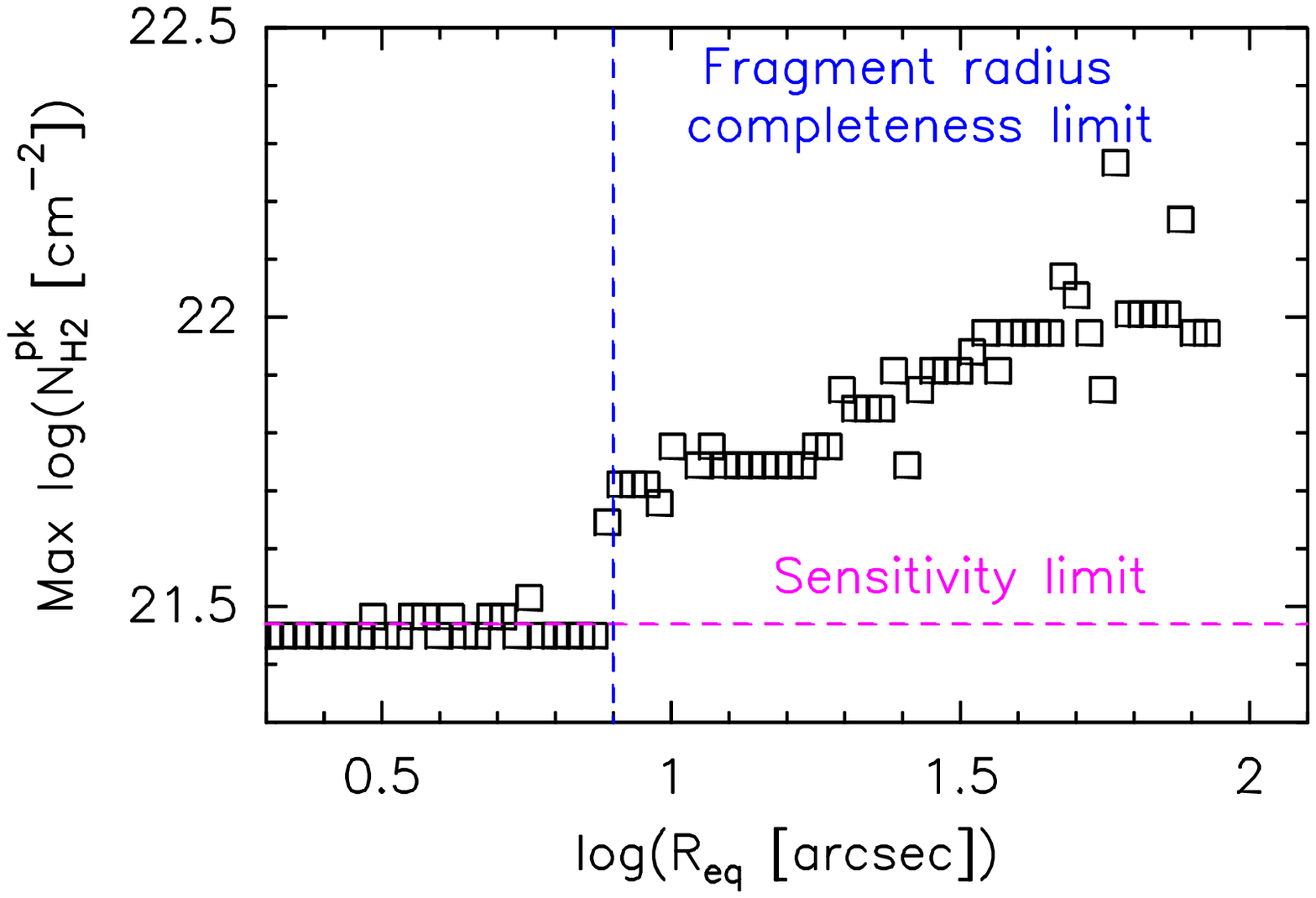}
\caption{Maximum peak column density for all IRDCs (left) and fragments (right) falling in a given bin of radius. We can see
that this maximum differs from the sensitivity limit $N_{\rm H_2}^{amp}$ when reaching a certain radius. This radius is taken as being the completeness radius.
\label{rad_comp}}
\end{figure*}

\section{Completeness limits}
\label{sec:complete}

\subsection{Column Density and Mass}

The mass completeness limit for the IRDCs and fragments can be written as 
\begin{equation}
M_{\rm c}= \pi(R_{\rm c} d_{\rm c})^2 \times <N_{\rm H_2}>_{\rm c} 
\label{eqn:mass}
\end{equation}
where $R_{\rm c}$ is the smallest radius above which the sample is complete,
$d_{\rm c}$ is the distance within which the majority of the sources occur,
and $<N_{\rm H_2}>_{\rm c}$ is the typical average column density of the
structures with a radius $R_{\rm c}$.
Figure~\ref{dist}(left) shows that about 95$\%$ of the IRDCs in that plot have
distances below 6~kpc and so we conservatively adopt $d_{\rm c}=6$~kpc.

Estimating $R_{\rm c}$ is less straighforward.  The completeness limits of our
survey are related to two parameters of the source extraction: $N_{\rm
  H_2}^{\rm amp}$, the minimum column density amplitude of a source (which is
related to sensitivity) from the boundary of a cloud to its peak; and the
angular resolution, 4\arcsec\ both for IRDCs and fragments. In order to
investigate how these contribute to the completeness limits we look at the
distribution of average column density of IRDCs for objects of a given range of
sizes as plotted in Fig.~\ref{pk_r}. We then plot the column density at the
peak of these distributions as a function of cloud size.  This is also done
for the fragments. Figure \ref{rad_comp} shows these plots.


The plots show a similar structure for both the IRDCs and fragments. Up to some
size, $55\arcsec$ for the IRDCs and $9\arcsec$ for the fragments, the peak of
the column density distributions is constant. Above these values it increases
with increasing size. This constant column density for small sizescales
suggests that the sample is not fully probing the populations of objects at
these sizescales. There are objects in these size ranges which have lower
column densities and are not sampled by the objects in the catalogue.  These
plots therefore show the size limit completeness of the catalogue, for both
IRDCs, $R_c=55\arcsec$, and fragments, $R_c=9\arcsec$.

Adopting these sizes, the average column density of
clouds/fragments below these sizes gives average column density
completeness limits of $ <N_{H_2}>_{\rm c}^{\rm frag}=
2.5\times10^{21}$~cm$^{-2}$ and $ <N_{H_2}>_{\rm c}^{\rm IRDC}=5\times10^{21}$~cm$^{-2}$. Using Eq.~\ref{eqn:mass}, these
values give the mass completeness of the catalogue as $M_{\rm c}^{\rm
  IRDC}= 800$~M$_{\odot}$ and $M_{\rm c}^{\rm frag}=9$~M$_{\odot}$.

\subsection{Density}

%
%
%
 
The density distribution of IRDCs is difficult to interpret since the
clouds are defined based on a column density threshold. Therefore we
confine our discussion of the density distributions to the fragments.
Both sensitivity and angular resolution are important limiting factors
in the context of density completeness of the sample: both compact
low-mass fragments and large, diffuse, high-mass fragments could remain
undetected.  For any fragment mass, $M_{\rm lim}$ there is a minimum
radius for which the peak column density of the fragment becomes
higher than the threshold for identifying fragments
($3\times10^{21}$cm$^{-2}$).  If this minimum radius is larger than the
angular resolution then such a fragment is detected. Therefore, we can
define a density completeness limit for all fragments more massive
than $M_{\rm lim}$. The minimum radius, $R_{\rm min}$, and the
corresponding maximum density, $\rho^{\rm low}_{\rm c}$, are given by
\begin{eqnarray}
  R_{\rm min} &=& \sqrt{\frac{M_{\rm lim}}{\pi \mu <N_{\rm H_2}^{\rm amp}>}}\\
  \rho_{\rm c}^{low} &=& \frac{3}{4}\sqrt{\frac{\pi}{M_{\rm lim}}} \mu^{3/2}<N_{\rm H_2}^{\rm amp}>^{3/2}
\end{eqnarray}
where $\mu$ is the mean mass per molecule and we adopt a column
density $ <N_{\rm H_2}^{\rm amp}>$ which is the average column density
of fragments of mass $M_{\rm lim}$ (which by definition have peak
column densities greater than the threshold to be identified as
fragments).
At a distance of 6~kpc, for fragment mass of 2/8/32~M$_{\odot}$ and $
<N_{\rm H_2}^{amp}> =1.5\times10^{21}$~cm$^{-2}$ we get R$_{\rm min}=5/10/20$\arcsec\
and $\rho_{\rm c}^{\rm low} =2.4/1.2/0.6\times10^3$cm$^{-3}$, respectively.

For the same given mass there is also an upper density limit which corresponds to the point where
the size of the fragment becomes smaller than the resolution of the observations.
This provides an upper limit on the density, $\rho_{\rm
  c}^{\rm up}=0.75\times M_{lim}/(\pi R_{res}^3)$. For the 3 masses discussed
before we get $\rho_{\rm c}^{up} =0.4/1.7/6.8\times10^5$~cm$^{-3}$.


\section{Size, mass and density structure of IRDCs}
\subsection{Physical size distribution}
\label{sec:size}

To calculate the density and mass of the IRDCs and fragments requires a
distance for each object.  Two different approaches to statistically attribute
a particular distance to a particular cloud have been adopted.  The first is
to simply assign a unique distance to all clouds. Doing this, the physical
size distribution of IRDCs and fragments will be exactly the same as the
angular size distribution.  Given the well peaked distribution of distances
for clouds with measured distances (Fig.~\ref{dist}), this should be a
reasonable first approximation. However, a more sophisticated approach is to
make use of the distribution of distances (rather than just its peak
position).  To do this we adopt a distance distribution for the IRDCs and then
randomly assign a distance drawn from this distribution to each cloud.  Doing
this for the whole sample of clouds repeatedly provides a statistical sampling
of the distance distribution.  The final physical size distribution is the
convolution of the true physical size distribution by the chosen distance
distribution. However this does not have a crucial impact on the
interpretation of the physical size distribution if the dispersion of the
distance distribution is much smaller than the angular size distribution. This
is clearly the case since the angular sizes, both for IRDCs and fragments,
extends over 2 order of magnitudes while IRDC distances span only over a
factor of 3 at most. In other words, the dispersion in distance has relatively
little effect on the final physical size distribution (see Appendix A).
To assign distances to the clouds using this sampling technique we adopt a
Gaussian distribution of distances with a peak at 4 kpc and a
dispersion of 1 kpc, consistent with observed distance distributions (Fig.~\ref{dist}).

\subsection{Mass distributions}

\begin{figure*}[!t!]
\hspace{0.5cm}
\vspace{-.cm}
\includegraphics[width=8cm,angle=270]{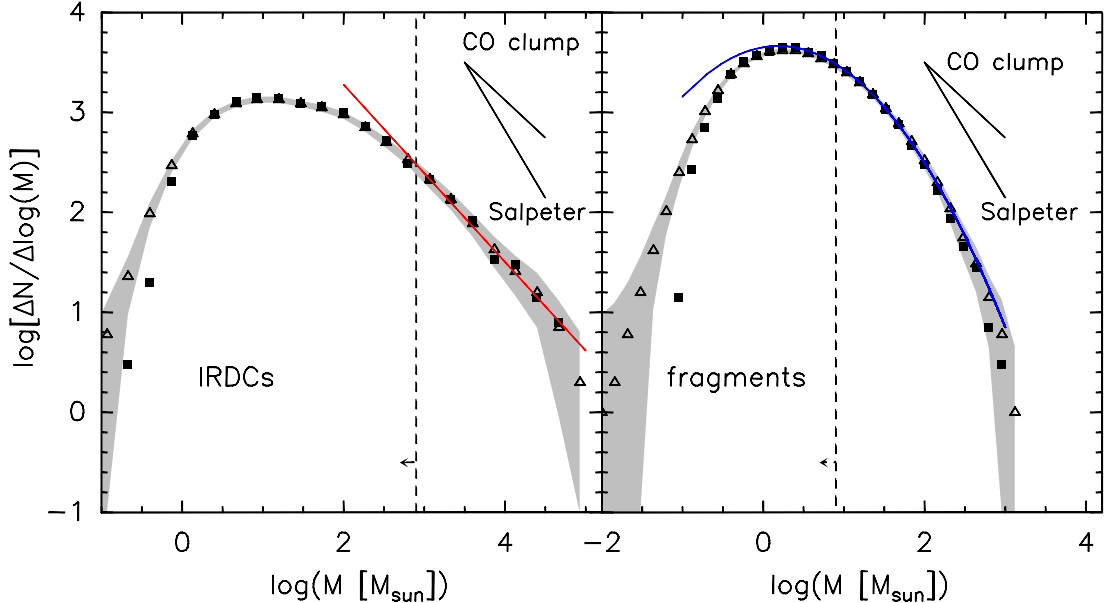}
\caption{ Mass distribution of infrared dark clouds (left) and
  fragments (right) for two different distance distributions: The
  filled square symbols correspond to the adoption of an unique
  distance of 4~kpc for each single cloud, while the open triangles
  and associated shaded area corresponds to adopting and sampling a
  Gaussian distance distribution (see Section 5.1).  The vertical
  dashed lines show the incompleteness limits. The best-fit is linear
  for the IRDCs (red solid line) with $\Delta N_{\rm IRDC}/\Delta
  \log(M) = M^{-\alpha}$ with $\alpha=0.85\pm0.07$, while the best-fit
  for fragments (blue solid line) is a lognormal function.  For
  comparison the slope of the mass distribution of CO molecular clouds
  and clumps and the Salpeter part of the IMF are also shown.}
 \label{mass}
\end{figure*}

Mass distributions of molecular cloud structures have been extensively studied
in the past, therefore they represent a good point of comparison for this
current study. We defined mass as:
\begin{equation}
M=\pi R_{\rm eq}^2<N_{\rm H_2}>
\end{equation}
where $<N_{\rm H_2}>$ is the average column density across the IRDC or
fragment and $R_{\rm eq}$ its equivalent radius (PF09).

Figure~\ref{mass} shows the mass distributions for IRDCs and fragments
calculated adopting a single distance of 4kpc (filled square symbols) and for
randomly attributed distances as described in Section~\ref{sec:size}.  The
shaded band on the figures shows the range (3 times the dispersion) spanned by
the 100 different distance realizations and the open triangles the mean for
the different realizations.  The completeness limits are shown by the dashed
lines.  For comparison the power-law slopes of the CO clump mass function
(slope $= -0.7$) and the Salpeter mass function (slope $= -1.35$) are also
shown.  Using the MPFITS IDL package (Markwardt 2009) we have fitted the two
distributions above their respective completeness limits.  For the IRDCs we
find a linear function (in a log-log plot) provides a good fit with
$dN_{\rm{IRDC}}/d\log M=M^{-\alpha}$ with $\alpha=0.85 \pm 0.07$.  The mass
distribution of fragments is better fitted by a lognormal function defined
as \begin{equation} \frac{dN_{\rm{IRDC}}}{d\log M}= A \exp( -
  [\log(M)-\log(M_{\rm{peak}})]^2/2\sigma^2)
\end{equation}
where A is a normalization constant, $M_{\rm{peak}}$ is the peak mass of the
distribution, and $\sigma$ is the dispersion. However, since we do not map the
peak, the precise parameters of the lognormal function fit are not well
constrained, several provide adequate fits to the data points. The function
showed in Fig.~\ref{mass} has A=4610, M$_{peak} =1.55$~M$_{\odot}$, and
$\sigma=0.78$. As argued above, the figure confirms that as a consequence of
the nature of the distance distribution, there is relatively little
difference in the derived mass distributions whether a single distance is
adopted for the clouds or statistical approach is adopted. Also, the results of this
analysis are also not strongly dependent on exact parameters of the assumed
distance distribution as demonstrated in Appendix~A.

A number of previous studies have attempted to construct, with samples at
least an order of magnitude smaller, the mass distributions of IRDCs (Simon et
al. 2006; Marshall et al. 2009) and fragments within them (Rathborne et
al. 2006; Ragan et al. 2009). Except for the Ragan et al. study, the mass
distributions in these studies agree: the IRDC mass distribution is similar to
that of CO clumps, while the distribution for the sub-structures are steeper,
more like the Salpeter IMF.

In their analysis of 11 IRDCs, Ragan et al. (2009) found that the mass
distributions of what they called {\it clumps}, which correspond to
fragments here, is quite flat, similar to the CO clump mass distribution, in
contrast with the present study.
However it is difficult to understand the Ragan et al. result as the radii and
masses they quote for their clumps imply 8\micron\ opacities over 10 times
larger than the 8\micron\ opacities they quote.

\subsection{Density distribution}

\begin{figure*}[!th!]
\hspace{0.5cm}
\includegraphics[width=6.5cm,angle=270]{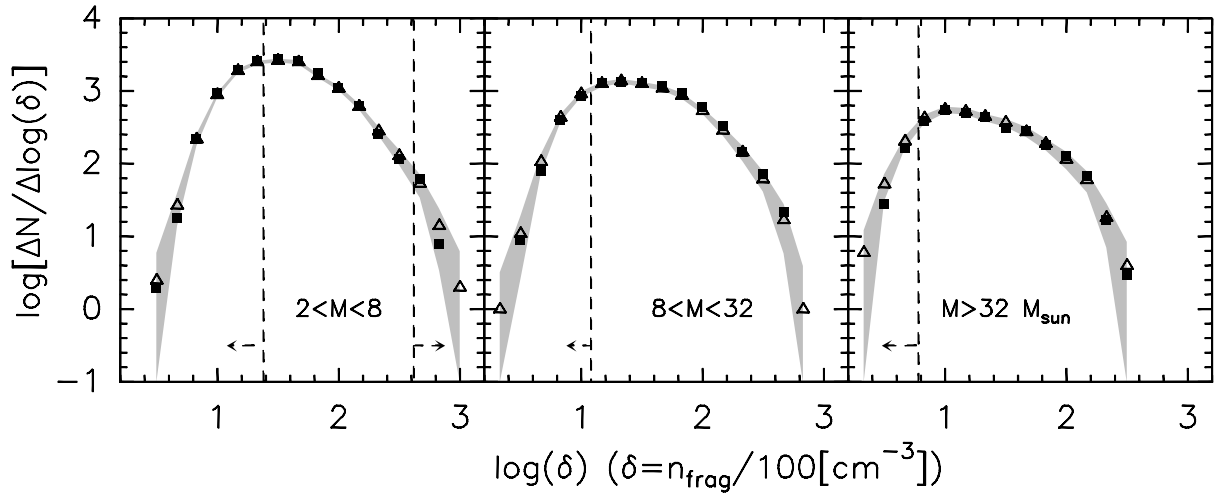}
\caption{Distribution of the number density of fragments normalised to 100
  cm$^{-3}$. {\it(left):} Fragments with a mass $2<M < 8$~M$_{\odot}$.
  {\it(middle):} Fragments with mass $8<M < 32$~M$_{\odot}$; {\it (right):}
  Fragments with a mass $> 32$~M$_{\odot}$. The dashed lines mark the density
  completeness limits, lower and upper for the left hand side panel and only
  lower for the 2 others (since the upper limits are off the plots). The
  square and triangle symbols and shaded area have the same meaning
  as for Fig.~\ref{mass}.
  \label{density}}
\end{figure*}

The density distribution of fragments may provide important insights on the
physical process generate these structures.  We define the number density of a
fragment as
\begin{equation}
n = <N_{H_2}>/d_t
\end{equation}
where d$_t$ is the line of sight size of the fragments which is assumed to be
twice the projected radius.  Figure \ref{density} shows the fragment density
distributions for the following mass ranges: $2<M<8$M$_\odot$,
$8<M<32$M$_\odot$ and $M>32$M$_{\odot}$.  For each range, the density
completeness limit, the dashed line, is calculated for $M_{\rm lim} =
2/8/32$~M$_{\odot}$, respectively. The density distribution over the entire
mass range is very similar to that for the lower mass range (left panel).  The
figures show that going from low mass to high mass fragments, the
distributions become flatter.  Compared to low density fragments, there is a
higher probability of finding high density fragments for high mass
fragments. One of the main issue in interpreting such a plot in terms of the
formation of the fragments is that the density of gravitationally bound
fragments is increasing over the time as they evolve (i.e. contract), and
therefore might {\it pollute} the initial density PDF of the parental
IRDCs. In the next section we will discuss the impact of such effects on the
density distributions.


\section{Discussion: Turbulent vs gravitational dominated structures}

The mass distributions of IRDCs and fragments plotted in Fig.~\ref{mass}
clearly shows a steepening, from large structures to smaller fragments. While
the mass distribution of IRDCs is similar to that of CO clumps, the fragment
mass distribution has a slope at high masses which is reminiscent of the slope
of the Salpeter IMF, although it is best fitted with a lognormal function. 
However, two biases could affect the shape of the high-mass end of the
distribution and the interpretation that its slope is related to the Salpeter
IMF. The detailed structure of this part of the distribution may be
particularly sensitive to the adopted Gaussian distance distribution. Also
high-mass fragments might evolve more rapidly than their low-mas analogues and
therefore be under represented in extinction observations at 8\microns
(c.f. Hatchell \& Fuller 2008).

\begin{figure}[!t!]
\hspace{0.cm}
\vspace{-.cm}
\includegraphics[width=5.5cm,angle=270]{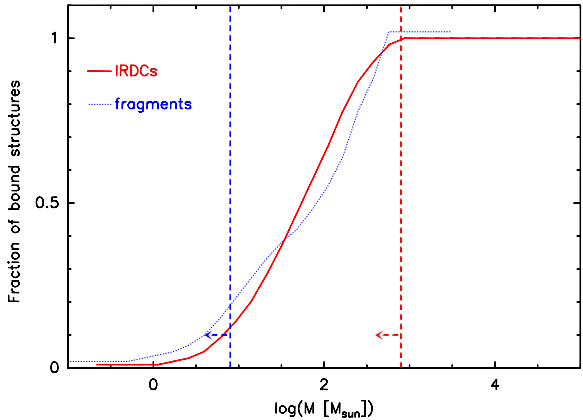}
\caption{Fraction of bound IRDCs and fragments as a function of their mass. The
  typical error due to distance uncertainty is 0.1. The dashed lines show the
  incompleteness limits of both fragments and IRDCs.
  \label{boundfrac}}%
\end{figure}

 To the first order, the mass distributions are
in agreement with the theoretical work of Hennebelle \& Chabrier (2008) who interpreted
the transition from a flat mass distribution to a steeper one 
 as the transition from turbulence-generated structures  to 
 gravity dominated structures. In this context it is interesting to measure the gravitational binding of IRDCs and fragments. To investigate this, we have used Larson's relation (Larson 1981) to compute the kinetic
support, and calculate the virial mass. Following Hennebelle \& Chabrier (2008) we assume the effective
velocity dispersion is given by
\begin{equation}
c_{\rm eff}= (c_s^2 +0.33V_0^2d^{2\eta})^{1/2}
\end{equation}
where $c_s$ is the sound speed $V_0 $ is the normalization velocity of
Larson's relation,  $\eta$ is the power law index of Larson's relation, and $d$ is the size of the structure. We can then use this to compute the corresponding virial mass M$_{\rm vir}$
 for every IRDC/fragment. Figure \ref{boundfrac} shows the fraction
of IRDCs and fragments having a ratio $M_{\rm H_2}/M_{\rm vir} > 0.5$,
assuming $c_s = 0.2$~km/s (T=10~K),  $V_0 = 1$~km/s, and $\eta=0.4$.  Above the
completeness limit all the IRDCs appear gravitationally bound, as do the
majority of the fragments.  Of course large uncertainties exist on the use of
Larson's relation and its normalization.  However, different normalizations
still give similar conclusions about the fraction of IRDCs and fragments which
are bound.

A consequence of the IRDCs being gravitationally bound is that the observed change of slope of the mass distributions shown in Fig.~\ref{mass}
does not represent the transition from turbulence dominated to gravity dominated structures: 
most of the IRDCs down to the completion limit are bound.
However, as shown in Fig.~\ref{boundfrac} a significant fraction of the fragments lying
above the completeness limit are unbound. 
 In other words, even if globally gravitationally bound, IRDCs may contain turbulence-generated over-densities which will probably disperse and not form stars. The physical properties of these unbound fragments are likely to be 
representative of the initial conditions of star formation within IRDCs.
%



\begin{figure*}[!th!]
\hspace{0.cm}
\vspace{-.cm}
\includegraphics[width=6.5cm,angle=270]{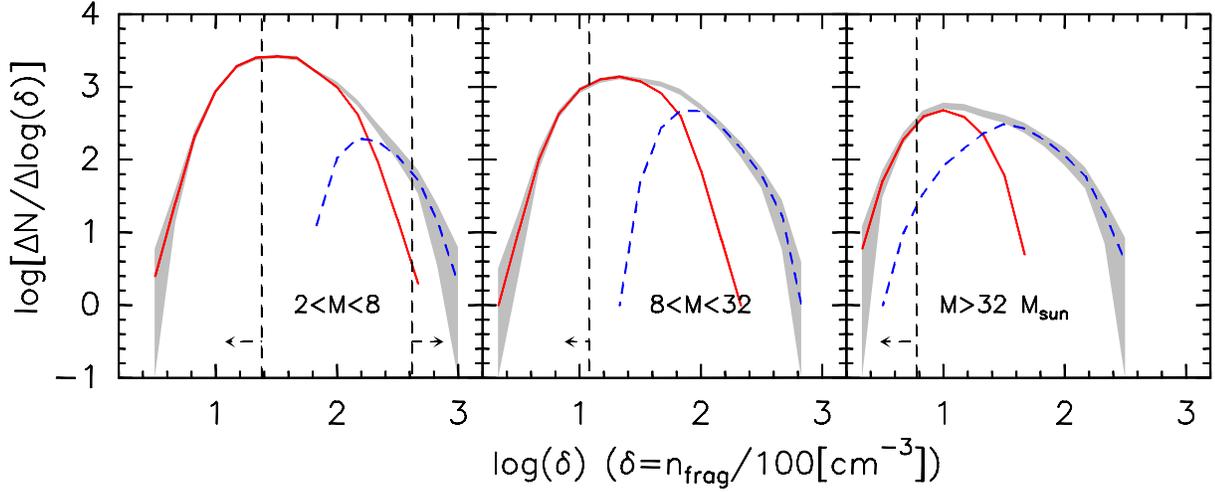}
\caption{Same as Fig.~\ref{density}, the shaded areas are the same as
  in Fig.~\ref{density}. The red solid line represent the density
  distributions of gravitationally unbound fragments, while the blue
  dashed line represent the gravitationally bound
  fragments.} \label{densbound}
\end{figure*}

Using the previously defined ratio to separate bound fragments to
unbound ones we constructed the density distributions for both type of
fragments as shown in Fig.~\ref{densbound}. The mass ranges are the
same as in Fig.~\ref{density}.  The unbound fragments have all very
similar distributions, independent of their mass range. In particular,
the high mass end of the distribution is well fit by the following
relation $\Delta N/ \Delta \log(n) \propto n^{-4.0\pm0.5}$, {\bf the error bar arising from 
the uncertainties on Larson's relation.} The location of
the peak seems to move with the completeness limit and is therefore
questionable. From these plots we cannot exclude a possible lognormal
distribution for unbound fragments,
 but the peak of such distribution would have to be below
n$\sim10^3$~cm$^{-3}$.

In contrast, the density distribution of the gravitationally bound
fragments show a well defined peak between n=$10^3$~cm$^{-3}$ and
$10^4$~cm$^{-3}$ in each mass range and a shape which broadens to
lower densities as the mass range increase. 
This could result from the higher mass fragments of a given density
evolving to being bound more rapidly that lower mass fragments.
%

\section{Summary}

We used the largest sample of IRDC column density maps to date in
order to better characterize the size, mass, and density structure of
dense molecular clouds. The large number of objects, 11,000 IRDCs and
50,000 fragments, allows a detailed analysis of the completeness of
the sample. Using a statistically attribute distances to each IRDC, we
have demonstrated that above the completeness limit the mass
distribution of the IRDCs are consistent with a power-law $\Delta
N_{\rm IRDC}/\Delta \log(M) = M^{-0.8}$, where $N_{\rm IRDC}$ is the
number of clouds. For the fragments the high mass end of the mass
distribution shows a steeper slope, consistent with the slope of the
Salpeter IMF, with the overall distribution well fitted by a lognormal
function.

Using Larson's law to estimate the linewidth of each IRDC and
fragment, we have shown that above our completeness limit all the
IRDCs and the majority of fragments are likely to be bound. This
implies that the transition in the shape of the mass distribution does
not reflect a transition from unbound to graviationally bound
structures. Looking at the distribution of fragment density shows that
bound fragments dominate the high density ($n\gtrsim10^4$~cm$^{-3}$)
end of the distribution for all mass ranges, and dominate the whole
distribution for the highest range of fragment masses.  There is also
a distinct broadening of the distribution with increasing fragment
mass. This could be a result of the higher mass fragments evolving to
being bound more rapidly that lower mass fragments.  The number of
unbound fragments as a function of number density has the form $\Delta
N_{\rm f}/ \Delta \log(n) \propto n^{-4.0\pm0.5}$ (where $N_{\rm f}$ is the
number of fragments) down to a density of $\sim10^3$~cm$^{-3}$ where
the completeness limit is reached.

The absence of bright infrared sources embedded in IRDCs indicates that the mass distributions and density distributions as a function of mass
and degree of gravitational binding derived here are representative of the initial conditions of star formation within dense molecular clouds. These results should serve as
constraints on theoretical and numerical models in order to identify and
characterize the physical processes responsible for the formation and
early fragmentation of molecular clouds. 

\begin{acknowledgements}
{\it Acknowledgements. We thank the anonymous referee and John Scalo for their thorough  reports which helped significantly improve the initial version of the paper. We would also like to thank Patrick Hennebelle for some useful discussions.}
\end{acknowledgements}

\appendix

\section{Different distance distributions}

\begin{figure*}[!th!]
\hspace{-0.cm}
\vspace{-.0cm}
\includegraphics[width=12cm,angle=270]{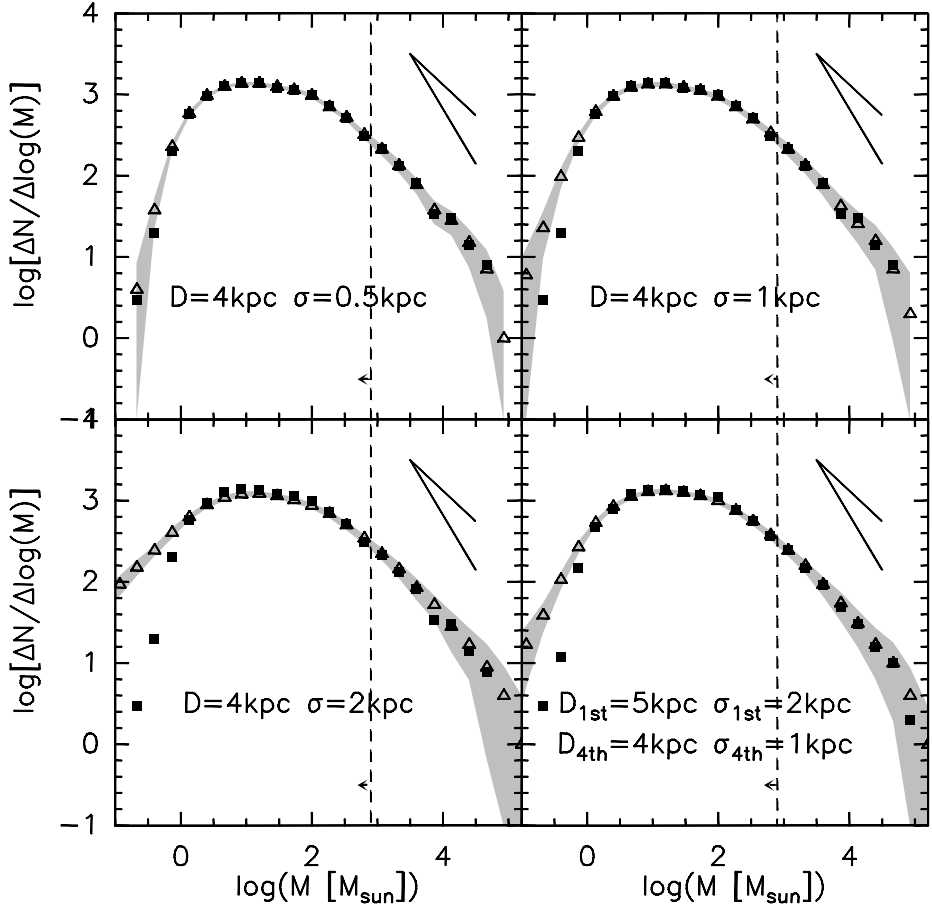}
\caption{Mass distribution of IRDCs. The different symbols have the
  same meaning as in Fig.~\ref{mass}. The four plots corresponds to
  four different assumptions on the IRDC distance distributions:
  (top-left) Gaussian distribution with a peak at 4kpc and a
  dispersion of 0.5kpc; (top right) Gaussian distribution with a peak
  at 4kpc and a dispersion of 1kpc; (bottom-left) Gaussian
  distribution with a peak at 4kpc and a dispersion of 2kpc;
  (bottom-right) Gaussian distribution with a peak at 5kpc and a
  dispersion of 2kpc for the 1st quadrant clouds and a Gaussian
  distribution with a peak at 4kpc and a dispersion of 1kpc for the
  4th quadrant clouds. Note that we also performed tests with a peak
  distance at 5kpc for all IRDCs but these resulted in distributions
  very similar to the 4kpc results shown here, but shifting the
  distributions slightly to more massive objects.  }\label{irdcpar}
\end{figure*}

\begin{figure*}[!th!]
\hspace{-0.cm}
\vspace{-.0cm}
\includegraphics[width=12cm,angle=270]{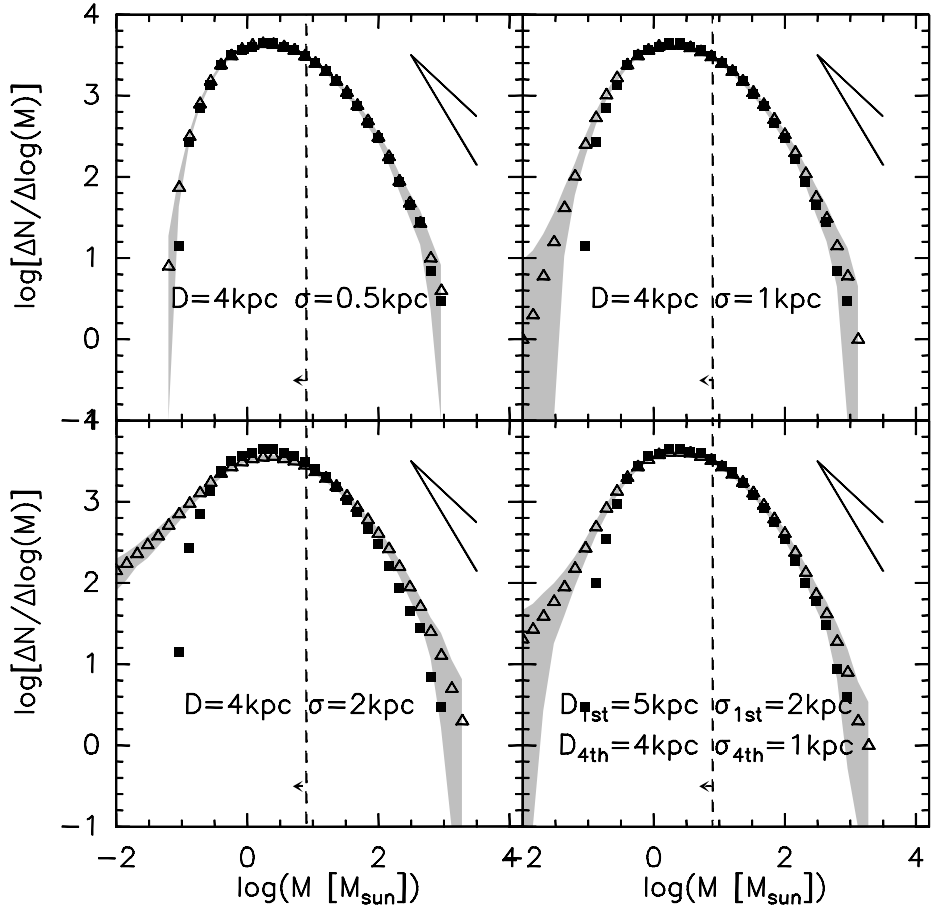}
\caption{Same as Fig.~\ref{irdcpar} but for the fragments. 
  \label{fragpar}}
\end{figure*}

In order to investigate further the impact of choosing a given
distance distribution on the calculated mass distributions we
performed a series of tests with different assumptions. Using the same
approach as described in Section 4 and 5, we used a Gaussian
distribution for distances with a mean distance of 4 and 5 kpc with
dispersions of 0.5, 1 and 2 kpc. We also performed a test assuming
different distance distributions for the 1st quadrant and 4th quadrant
IRDCs. We show the resulting mass distributions in Fig.~\ref{irdcpar}
and \ref{fragpar}. We see that the shape of the mass distributions
above the completeness limits for both the IRDCs and fragments, remain
basically the same, with best fit parameters changing only of a few
0.1 dex.

\end{document}